\documentclass[structstract]{aa}  
\usepackage{graphicx,longtable,cancel}
\usepackage{txfonts}
\begin{document}

   \title{Diversity of GRB energetics vs. SN homogeneity: supernova 2013cq associated with the gamma-ray burst 130427A\thanks{Based on observations made with the VLT, operated on the mountain of Cerro Paranal in Chile under program 091.D-0291(A) and with the TNG, operated on the island of La Palma by the Fundaci\'{o}n Galileo Galilei of the INAF (Instituto Nazionale di Astrofisica) at the Spanish Observatorio del Roque de los Muchachos of the Instituto de Astrof\'{i}sica de Canarias under program A27TAC$\_$5.}}


   \author{A. Melandri\inst{1}, E. Pian\inst{2,3,4}, V. D'Elia\inst{5,6}, P. D'Avanzo\inst{1},  M. Della Valle\inst{7,8}, P. A. Mazzali\inst{9,10,11}, G. Tagliaferri\inst{1}, Z. Cano\inst{12},  A. J. Levan\inst{13}, P. M\cancel{o}ller\inst{14},
   L. Amati\inst{3}, M. G. Bernardini\inst{1}, D. Bersier\inst{9}, F. Bufano\inst{15}, S. Campana\inst{1}, A. J. Castro-Tirado\inst{16}, S. Covino\inst{1}, G. Ghirlanda\inst{1}, K. Hurley\inst{17}, D. Malesani\inst{18}, N. Masetti\inst{3},  E. Palazzi\inst{3}, S. Piranomonte\inst{6}, A. Rossi\inst{19}, R. Salvaterra\inst{20}, R. L. C. Starling\inst{21}, M. Tanaka\inst{22}, N. R. Tanvir\inst{21}, S. D. Vergani\inst{23}}

   \institute{
   $^{1}$ INAF - Osservatorio Astronomico di Brera , via E. Bianchi 46, I-23807, Merate (LC), Italy\\
              \email{andrea.melandri@brera.inaf.it}\\
   $^{2}$ Scuola Normale Superiore, Piazza dei Cavalieri 7, I-56126 Pisa, Italy\\
   $^{3}$ INAF - Istituto di Astrofisica Spaziale e Fisica Cosmica, Via P. Gobetti 101, I-40129 Bologna, Italy\\
   $^{4}$ INFN, Sezione di Pisa, Largo Pontecorvo 3, I-56127 Pisa, Italy\\
   $^{5}$ ASI, Science Data Centre, Via del Politecnico snc, I-00133 Roma, Italy\\
   $^{6}$ INAF - Osservatorio Astronomico di Roma, Via Frascati 33, I-00040, Monte Porzio Catone (RM), Italy\\
   $^{7}$  INAF - Osservatorio Astronomico di Capodimonte, Salita Moiariello 16, I-80131 Napoli, Italy\\
   $^{8}$ ICRANET, Piazza della Repubblica 10, I-65122 Pescara, Italy\\
   $^{9}$ ARI, Liverpool John Moores University, IC2 Liverpool Science Park 146 Brownlow Hill, Liverpool, L3 5RF, UK\\
   $^{10}$ INAF - Osservatorio Astronomico di Padova, Vicolo dell'Osservatorio 5, I-35122 Padova, Italy\\
   $^{11}$ Max-Planck Institute for Astrophysics, Garching, Karl-Schwarzschild-Str. 1, Postfach 1317, D-85741 Garching, Germany\\
   $^{12}$ Centre of Astrophysics and Cosmology, Science Institute, University of Iceland, Dunhagi 5, IS-107, Reykjavik, Iceland\\
   $^{13}$ Department of Physics, University of Warwick, Coventry, CV4 7AL, UK\\
   $^{14}$ European Southern Observatory, Karl-Schwarzschildstrasse 2, D-85748 Garching bei M\"{u}nchen, Germany\\
   $^{15}$ Departamento de Ciencias Fisicas, Universidad Andres Bello, Avda. Republica 252, Santiago, Chile\\
   $^{16}$ Instituto de Astrof\'{i}sica de Andaluc\'{i}a, Glorieta de la Astronom\'{i}a s/n, 18008, Granada, Spain\\
   $^{17}$ University of California, Berkeley, Space Sciences Laboratory, 7 Gauss Way, Berkeley, CA 94720-7450, USA\\
   $^{18}$ Dark Cosmology Centre, Niels Bohr Institute, University of Copenhagen, Juliane Maries Vej 30, 2100, Copenhagen, Denmark\\
   $^{19}$ Th\"{u}ringer Landessternwarte Tautenburg, Sternwarte 5, 07778, Tautenburg, Germany\\
   $^{20}$ INAF - Istituto di Astrofisica Spaziale e Fisica Cosmica Milano, via E. Bassini 15, I-20133 Milano, Italy\\
   $^{21}$ Department of Physics and Astronomy, University of Leicester, University Road, Leicester LE1 7RH, UK\\
   $^{22}$ National Astronomical Observatory of Japan, Mitaka, Tokyo 181-8588, Japan\\
   $^{23}$ GEPI, Observatoire de Paris, CNRS, Univ. Paris Diderot, 5 Place Jules Jannsen, F-92195, Meudon, France
             }

   \date{}

\abstract
   {}
   {Long-duration gamma-ray bursts (GRBs) have been found to be associated with broad-lined type-Ic supernovae (SNe), but only a handful of cases have been studied in detail. Prompted by the discovery of the exceptionally bright, nearby GRB\,130427A (redshift $z$ = 0.3399), we aim at characterising the properties of its associated SN\,2013cq. This is the first opportunity to test directly the progenitors of high-luminosity GRBs.}
   {We monitored the field of the \textit{Swift} long duration GRB\,130427A using the 3.6-m TNG and the 8.2-m VLT  during the time interval between 3.6 and 51.6 days after the burst. Photometric and spectroscopic observations revealed the presence of the  type Ic SN\,2013cq.}
   {Spectroscopic analysis suggests that SN\,2013cq resembles two previous GRB-SNe, SN\,1998bw and SN\,2010bh associated with GRB\,980425 and XRF\,100316D, respectively. The bolometric light curve of SN\,2013cq, which  is significantly affected by the host galaxy contribution,  is systematically more luminous than that of SN\,2010bh ($\sim$2 mag at peak), but is consistent with SN\,1998bw. The comparison with the light curve model of another GRB-connected SN\,2003dh, indicates that SN\,2013cq is consistent with the model when brightened by 20\%. This suggests a synthesised radioactive $^{56}$Ni mass of $\sim 0.4 M_\odot$. GRB\,130427A/SN\,2013cq is the first case of low-$z$ GRB-SN connection where the GRB energetics are extreme ($E_{\rm \gamma , iso} \sim 10^{54}$~erg).    We show that the maximum luminosities attained by SNe associated with GRBs span a very narrow range, but those associated with XRFs are significantly less luminous.  On the other hand the isotropic energies of the accompanying GRBs span 6 orders of magnitude (10$^{48}$~erg~$< E_{\rm \gamma , iso} <$ 10$^{54}$~erg), although this range is reduced when corrected for jet collimation. The GRB total radiated energy is in fact a small fraction of the SN energy budget.}
   {}


\authorrunning{A. Melandri et al.}
\titlerunning{Diversity of GRB energetics vs. SN homogeneity}

   \maketitle

%

\section{Introduction}

The majority of nearby under-energetic long duration gamma-ray bursts and X-ray flashes (GRBs and XRFs\footnote{XRFs are a softer version of GRBs, having integrated spectra peaking around 5-10 keV instead of 100-1000 keV.} with total isotropic energy $E_{\rm iso} <$ 10$^{51}$~erg) are associated with with highly energetic SNe (\cite{mazzali6}; \cite{mazzali5}). The study of the supernova in these cases is facilitated by the weakness of the GRB afterglow and the relatively small distance. 

For $z \lesssim$ 0.3 the accuracy of optical photometric and spectroscopic monitoring is satisfactory and individual atomic species can be identified in the SN spectrum (\cite{galama}; \cite{patat}; \cite{hjorth}; \cite{stanek}; \cite{soderberg2}; \cite{malesani}; \cite{ferrero}; \cite{pian}; \cite{sollerman}; \cite{mirabal}; \cite{bersier}; \cite{cano}; \cite{bufano}; \cite{mela}; \cite{schulze}). At higher redshifts, up to $z \sim$ 1, the presence of an associated SN is inferred from the detection of a ``rebrightening" in the late afterglow light curve due to the emerging SN component (\cite{bloom}; \cite{ctg1}; \cite{galama2}; \cite{ctg2}; \cite{lazzati};  \cite{masetti}; \cite{zeh}; \cite{gorosabel}; \cite{bersier}; \cite{cobb}; \cite{soderberg}; \cite{tanvir}; \cite{cano2}). In a few cases, single epoch spectra obtained close to the peak of this rebrightening show SN features similar to those exhibited by GRB-accompanying SNe at lower redshifts (\cite{max0}; \cite{garnavich}; \cite{greiner}; \cite{soderberg0}; \cite{berger}; \cite{max1}; \cite{sparre}; \cite{jin}) then confirming the association of the GRB with a SN event.

All GRBs associated with SNe at $z \lesssim$ 0.3 so far discovered show low isotropic energy, typically less than $\sim 10^{50}$ erg, with GRB\,030329/SN\,2003dh (\cite{hjorth}; \cite{stanek}; \cite{matheson}) being the only exception. This is a relatively nearby GRB ($z$ = 0.168) with an isotropic energy of $E_{\rm \gamma , iso} \sim 2 \times 10^{52}$~erg (\cite{spek}) which falls in the faint tail of the ``cosmological" GRBs energy distribution (e.g. \cite{amati2}; \cite{amati}). The long nearby GRB\,130427A is exceptional and outstanding. It showed a huge isotropic energy ($E_{\rm iso} \sim 10^{54}$~erg; \cite{maselli2}; \cite{amati5}) and at the same time the association with a SN was clear (SN\,2013cq; \cite{deuga}). GRB\,130427A follows the well-known Amati ($E_{\rm peak}$-$E_{\rm iso}$, \cite{amati2}) and Yonetoku ($E_{\rm peak}$-$L_{\rm iso}$, \cite{yonetoku}) correlations (\cite{maselli2}). This made the study of the properties and evolution of SN\,2013cq particularly interesting, since no very energetic GRB has ever been detected at relatively low redshift, so that this is the first occurrence of connection between a SN and a GRB that has all the characteristics of a cosmological event.

In this paper we present the results of our photometric and spectroscopic campaign, covering $\sim$ 1.5 months, carried out with the VLT and the TNG. Throughout the paper we assume a standard cosmology with $H_{\rm 0}$ = 72 km s$^{-1}$ Mpc$^{-1}$, $\Omega_{\rm m}$ = 0.27, and $\Omega_{\rm \Lambda}$ = 0.73.

\section{GRB\,130427A / SN\,2013cq}

GRB\,130427A was a long and extremely bright GRB ($T_{\rm 90} \sim 160$~s; \cite{scott}) that triggered independently the {\it Fermi} satellite at 07$^{\rm h}$ 47$^{\rm m}$ 06$^{\rm s}$.42 UT (\cite{vonkie}; \cite{zhu1}) and the {\it Swift} satellite at 07$^{\rm h}$ 47$^{\rm m}$ 57$^{\rm s}$.5 UT (\cite{maselli1}), and attained the highest fluences observed in the $\gamma$-ray band for both satellites ($f_{\it Swift}$ (15-150 keV) $\sim 5 \times 10^{-4}$~erg~cm$^{-2}$ and $f_{\it Fermi}$ (0.01-20 MeV) $\sim 4 \times 10^{-3}$~erg~cm$^{-2}$). The high-energy emission of GRB\,130427A was also detected by several orbiting observatory, i.e. MAXI (\cite{kawamuro}), INTEGRAL (\cite{pozanenko}), Konus-Wind (\cite{golenetskii}), AGILE (\cite{verrecchia}), Suzaku (\cite{akiyama}), RHESSI (\cite{smith}) and Mars Odyssey\footnote{http://www.ssl.berkeley.edu/ipn3/masterli.txt}. This event has been the focus of several recently published studies (\cite{zhu2}; \cite{preece}; \cite{maselli2}; \cite{vestrand}; \cite{kouveliotou}; \cite{laskar}; \cite{perley}; \cite{pana}; \cite{mg}). The redshift was measured to be $z$ = 0.3399 $\pm$ 0.0002 (\cite{levan1}; \cite{xu1}; \cite{flores}).

In the optical band a bright flash, probably due to a reverse shock component, was observed simultaneously with the high energy emission ($> 100$~MeV) at very early times. Subsequently, the GRB afterglow emission can be described by the contribution of both reverse and forward shocks (\cite{vestrand}; \cite{pana}; \cite{maselli2}; \cite{perley}; \cite{levan2}). 

GRB\,130427A exploded in a relatively bright, extended host galaxy, catalogued in the Sloan Digital Sky Survey (SDSS J113232.84+274155.4), that showed the typical properties of the nearby GRB host population (\cite{savaglio}). Its stellar mass ($M_{*} = 2.1 \pm 0.7 \times 10^{9}$ $M_{\odot}$) and mean population age ($\sim 250$ Myr) indicate a blue, young and low-mass galaxy (\cite{perley}). The afterglow of GRB\,130427A is slightly offset from the centroid of its host galaxy ($\sim 0.83^{\prime\prime}$, corresponding to $\sim 4$ kpc in projection at the redshift of the GRB) and apparently there is no highly star forming region underlying the GRB (\cite{levan2}).

Despite its relatively low redshift, which favoured the detection and follow-up of the associated SN\,2013cq in the $R$-band (\cite{xu2}), GRB\,130427A displayed all properties of more commonly observed high-redshift bursts. The extraordinarily  high  observed energetics of GRB\,130427A and its closeness motivated our optical multi-band search and intensive follow-up of its associated supernova.

\section{Observations and data reduction}

We observed the field of SN\,2013cq with the ESO 8.2-m Very Large Telescope (VLT) at Paranal Observatory  equipped with FORS2 (imaging in the $BVRI$ filters and spectroscopy) and with the Italian 3.6-m Telescopio Nazionale Galileo (TNG) equipped with DOLORES (imaging in the $g'r'i'$ filters) from 3.6 to 51.6 days after the burst. Tables \ref{logspec} and \ref{logmag} summarise our observations. 
 
\subsection{Imaging}

\begin{figure}
   \centering
      \includegraphics[width=7.0cm,height=8.8cm,angle=-90]{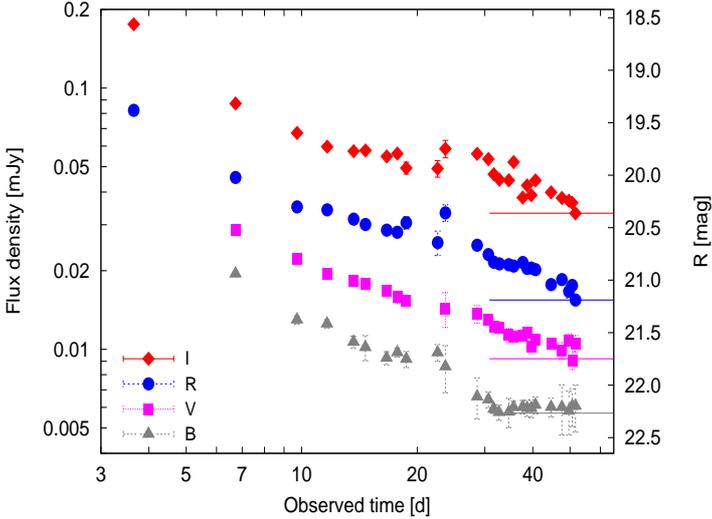}
      \caption{Observed $BVRI$ light curves of GRB\,130427A/SN\,2013cq, not corrected for Galactic extinction. For clarity, $R$ and $I$ magnitudes have been shifted by 0.45 and 0.75 mag, respectively. Horizontal lines represent the fluxes of the host galaxy in each filter.} 
    \label{lcSN}
\end{figure}

Image reduction, including de-biasing and flat-fielding, was carried out following standard procedures. Images were calibrated using a set of SDSS catalogued stars acquired with SDSS $g'r'i'$ filters (TNG observations) and with respect to standard fields in the $BVRI$ filters (VLT observations). We performed differential photometry at the position of the optical afterglow, using an aperture of $\geq$ 2 $\times$ FWHM of individual frames, large enough to include also the contribution of the underlying host galaxy. Observed $BVRI$ light curves are shown in Fig. \ref{lcSN}.

During our campaign it was not possible to acquire a deep image of the host galaxy alone because the GRB optical counterpart dominates the emission at early epochs and may still contribute significantly at late epochs.  Moreover the offset of the SN with respect to the host galaxy is only $\sim 0.8$\arcsec (\cite{levan2}), always comparable with the seeing of our images (see Table 3). Therefore, we used the SDSS magnitudes of the host galaxy that are $u = 22.41 \pm 0.33$, $g = 21.98 \pm 0.11$, $r = 21.26 \pm 0.09$, $i = 21.19 \pm 0.16$, $z = 21.11 \pm 0.54$. We converted these magnitudes into Johnson/Cousins magnitudes using the transformation equations in Jester et al. (2005), obtaining the final values $B_{\rm HG} = 22.2$, $V_{\rm HG} = 21.5$, $R_{\rm HG} = 21.2$, $I_{\rm HG} = 20.5$. These values are consistent with Perley et al. (2013) and shown in Fig. \ref{lcSN} as horizontal lines. 

We subtracted the estimated values for the host galaxy from our data, corrected for Galactic extinction (using the catalogued value of $E_{B-V,\,\mathrm{Galactic}}$ = 0.02 mag; \cite{schle}; \cite{schla}), applied k-corrections using our spectra (only $B$ and $V$ filters, because the spectra do not cover the redshifted $R$ and $I$ filters central wavelengths) and then subtracted the afterglow component. The temporal behaviour of the afterglow in each filter, where the early light curves were modelled with a forward relativistic shock into the circumstellar medium, can be described with a steepening power-law having decay indices of $\sim$ 0.8 and 1.5 before and after a break located at $\sim 0.5$ days (\cite{laskar}; \cite{maselli2}; \cite{perley}; \cite{xu2}). Our data cover the phase after this temporal break, and are consistent with the above time decay until a week after explosion, when the contribution from SN\,2013cq becomes increasingly important. The afterglow model was then subtracted from the host-subtracted light curves and the residual, which is attributed to SN\,2013cq only, was corrected for intrinsic absorption following Xu et al. (2013b). These authors estimated the value E$_{\rm (B-V),~Host~Galaxy}$ = 0.05 mag from the detection of Na I D 5890 $\&$ 5896 absorption lines. The final corrected $VRI$ light curves are reported in Fig. \ref{lc_rf_SN}. The $B$-band light curve is heavily contaminated by the host galaxy and therefore not meaningful and not shown. The de-reddened magnitudes have been transformed into monochromatic fluxes using the zeropoints in Fukugita et al. (1995).

\subsection{Spectroscopy}

 \begin{table}
 \caption[]{Journal of the VLT/FORS2 spectroscopic observations of SN\,2013cq. $\Delta t_{\rm obs}$ corresponds to the acquisition starting time of the spectra in the observer frame, with respect to the GRB trigger time, while $t_{\rm exp}$ is the total exposure. The $\langle$S/N$\rangle$ is the average per spectral bin between the S/N at the edges of the grism, where efficiency is lower, and that at the central wavelengths. The last column shows the rest-frame time.}
 \label{logspec}
 \centering
 \begin{tabular}{@{\hspace{5pt}}lccc@{\hspace{8pt}}c@{\hspace{8pt}}c}
 \hline
Date & $\Delta t_{\rm obs}$  & $t_{\rm exp}$ & $\langle$S/N$\rangle$ & Grism & $\Delta t_{\rm RF}$\\
\hline
 & (days) & (s) & & & (days)\\
\hline
2013 May 4 & 6.737 & 1$\times$900 & 13-28  & 300V & +5.02\\
2013 May 4 & 6.756 & 1$\times$900 & 2-6  & 300I & +5.04\\
2013 May 7 & 9.735 & 1$\times$900& 4-10 & 300V & +7.26\\
2013 May 10 & 13.669 & 1$\times$900 & 6-10 & 300I & +10.20\\
2013 May 12 & 15.660  & 2$\times$900 & 5-12 & 300V & +11.68\\
2013 May 14 & 16.680  & 2$\times$1800 & 10-20  & 300V & +12.44\\
2013 May 16 & 18.696 & 2$\times$900 & 3-9 &  300V & +13.95\\
2013 May 25 &  28.676 & 2$\times$900 & 3-5 &  300V & +21.40\\
2013 May 27 & 30.663  & 2$\times$1800 & 7-13  & 300V & +22.88\\
2013 May 28 & 31.665  & 2$\times$1800 & 10-15  & 300V & +23.63\\ 
\hline
\end{tabular}
\end{table}

VLT/FORS2 spectroscopy was carried out using the 300V grism, covering the range 4450--9500\,\AA~(corresponding to 3320--7090\,\AA~in the GRB rest-frame). For two epochs we also used the 300I grism, covering the range 6000--11000\,\AA~(4478--8210\,\AA~rest-frame). We used in all cases a $1''$ slit, resulting in an effective resolution $R = 440$ at the central wavelengths $\lambda_{\rm 300V} = 5900\,\AA$ and $\lambda_{\rm 300I} = 8600\,\AA$, respectively. The spectra were extracted using standard procedures within the packages ESO-MIDAS\footnote{\texttt{http://www.eso.org/projects/esomidas/}} and IRAF\footnote{\texttt{http://iraf.noao.edu/}}. A He-Ar lamp and spectrophotometric stars were used to calibrate the spectra in wavelength and flux, respectively. We accounted for slit losses by matching the flux-calibrated spectra to our simultaneous multi-band photometry.

After subtracting the contribution of the host galaxy using a spline  interpolation to the SDSS magnitudes, we corrected the residual for Galactic extinction (with the extinction curve of \cite{cardelli}). We also applied the correction for the intrinsic reddening following Xu et al. (2013b) as in section 3.1. We model the afterglow spectrum with a single power-law as
\begin{eqnarray}
F_\lambda(\lambda) = N \times \Bigg( \frac{\lambda_{\rm norm}}{\lambda} \Bigg)^{\beta_{\lambda}}
\end{eqnarray}
~\\
\noindent where we fixed $\lambda_{\rm norm}$ to the rest-frame flux at 6588 $\AA$ (corresponding to the $R$-band), $N$ is the power-law normalisation and the spectral index $\beta_{\lambda}$ = 1.5 (\cite{maselli2}, \cite{perley}). Telluric absorption features and the noisier parts of the spectra ($\lambda_{\rm RF} <$ 3400~$\AA$ and $\lambda_{\rm RF} >$ 6800~$\AA$ for the 300V grism and $\lambda_{\rm RF} <$ 4600~$\AA$ for the 300I grism) have been omitted. A clearly spurious feature has been deleted from the spectrum acquired at $+11.68$~rest-frame days.

\section{Results}

\begin{figure}
   \centering
   \includegraphics[width=8.8cm,height=7.0cm]{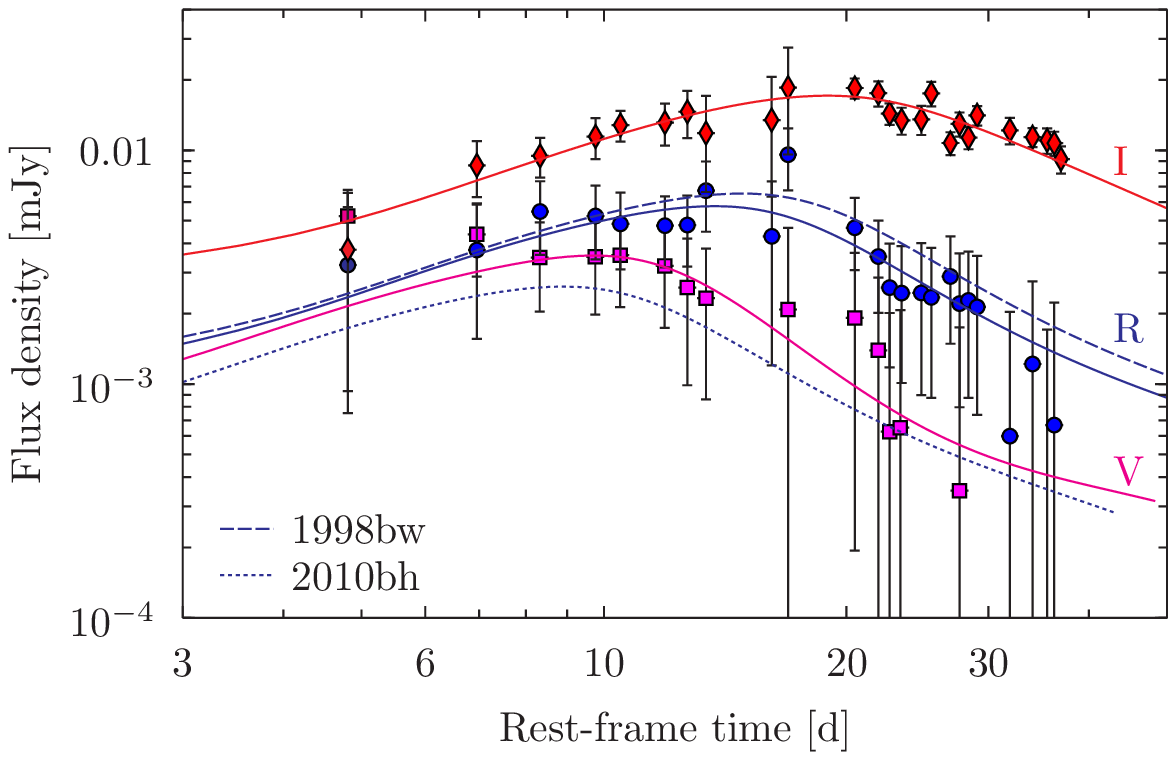}
      \caption{SN\,2013cq rest-frame light curve for the $V$ (magenta squares), $R$ (blue circles) and $I$ (red diamonds) bands. We show also the k-corrected $R$-band light curves of SN\,1998bw (dashed line) and SN\,2010bh (dotted line). Solid coloured lines are the k-corrected light curves of SN\,1998bw transformed by stretch and luminosity factors and represent the best fit to the SN\,2013cq $VRI$ light curves.}
    \label{lc_rf_SN}
\end{figure}

\subsection{Optical light curves}

Inspection of Fig. \ref{lcSN} shows the presence of a rebrightening at $\sim$ 20 days, more pronounced in the redder filters, that is the tell-tale signature of an underlying SN, first identified by Xu et al. (2013a) and named SN\,2013cq (\cite{deuga}). In Fig. \ref{lc_rf_SN} we show the final $VRI$ light curves of SN\,2013cq.

Also plotted in Fig. \ref{lc_rf_SN} are the k-corrected, $R$-band light curves of SN\,1998bw and SN\,2010bh, which were associated with GRB 980425 (\cite{galama}) and XRF\,100316D (\cite{starling}; \cite{cano}; \cite{bufano}; \cite{olivares}), respectively, as they would appear if they occurred at a redshift of $z$ = 0.3399. In order to determine the rest-frame peak times in each filter, we adopted the formalism of Cano (2013), by which the optical light curves of the well observed type Ic SN\,1998bw are used as templates to describe SNe with less well sampled light curves. The solid lines in Fig. \ref{lc_rf_SN} are the k-corrected light curves of SN\,1998bw in the relevant bands, stretched in time and flux to match SN\,2013cq. The scaling factors were obtained with a best-fit to the data of SN\,2013cq, following Cano (2013). The peaks of these template light curves represent our best estimates of the light maxima of SN\,2013cq: $T_{\rm peak, \it {V}}$ $\sim 9.6 \pm 0.7$, $T_{\rm peak, \it{R}}$ $\sim 13.8 \pm 0.9$ and $T_{\rm peak, \it{I}}$ $\sim 17.9 \pm 1.4$ days after the burst. We note that the errors are formal uncertainties returned by the fit and are likely to underestimate the real uncertainties by about a factor of 2. Our best-fit $R$-band maximum flux and time agree with those determined by Xu et al. (2013b) within the error bars. The SN\,2013cq flux maximum in the $R$-band is found to be slightly fainter ($\sim 0.2$~mag) than SN\,1998bw.

It is commonly observed in SNe of all types, including nearby Ic SNe and GRB/XRF SNe (\cite{richmond}; \cite{foley}; \cite{mazzali}; \cite{mazzali2}; \cite{valenti}; \cite{galama}; \cite{soderberg2}; \cite{soderberg};  \cite{bufano}), that the light maximum is reached later in redder bands.  In SN\,2013cq this temporal evolution is particularly fast, so that the rise in the $V$ band is rapid and resembles the one of XRF SNe (\cite{pian}; \cite{ferrero}; \cite{mirabal};  \cite{bufano}; \cite{olivares}), while in $R$ and $I$ bands the rise is more similar to that of SNe associated with low-$z$ under-luminous GRBs and classical GRBs at higher redshifts (\cite{galama}; \cite{patat}; \cite{garnavich}; \cite{malesani}; \cite{max2}; \cite{clocchiati}; \cite{mela}).


\subsection{Optical spectra}

\begin{figure*}
   \vspace{-5mm}
         \hspace{-10mm}
   \includegraphics[width=10.5cm,height=6.7cm]{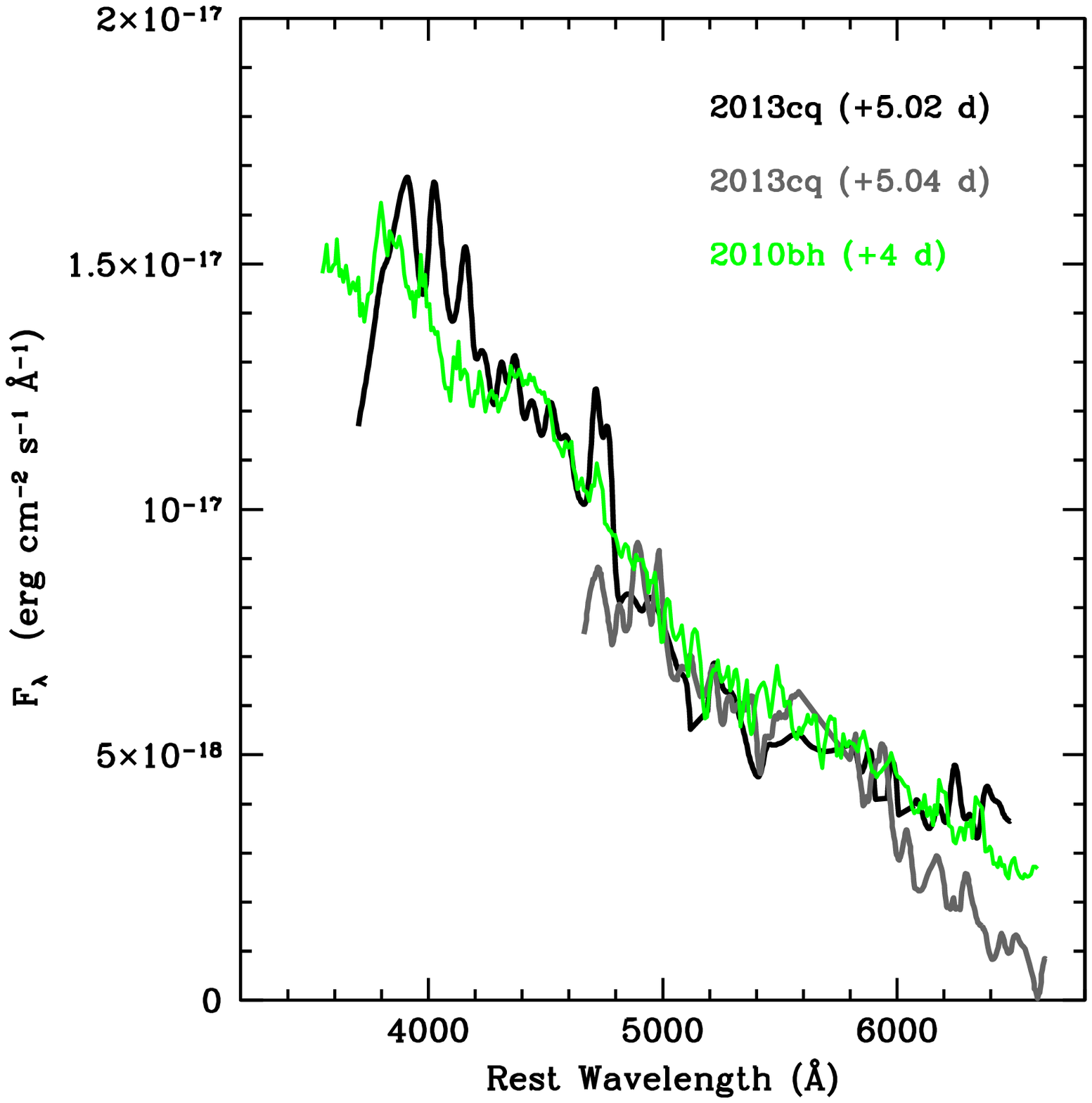}
   \vspace{-10mm}
         \hspace{-8mm}
   \includegraphics[width=10.5cm,height=6.7cm]{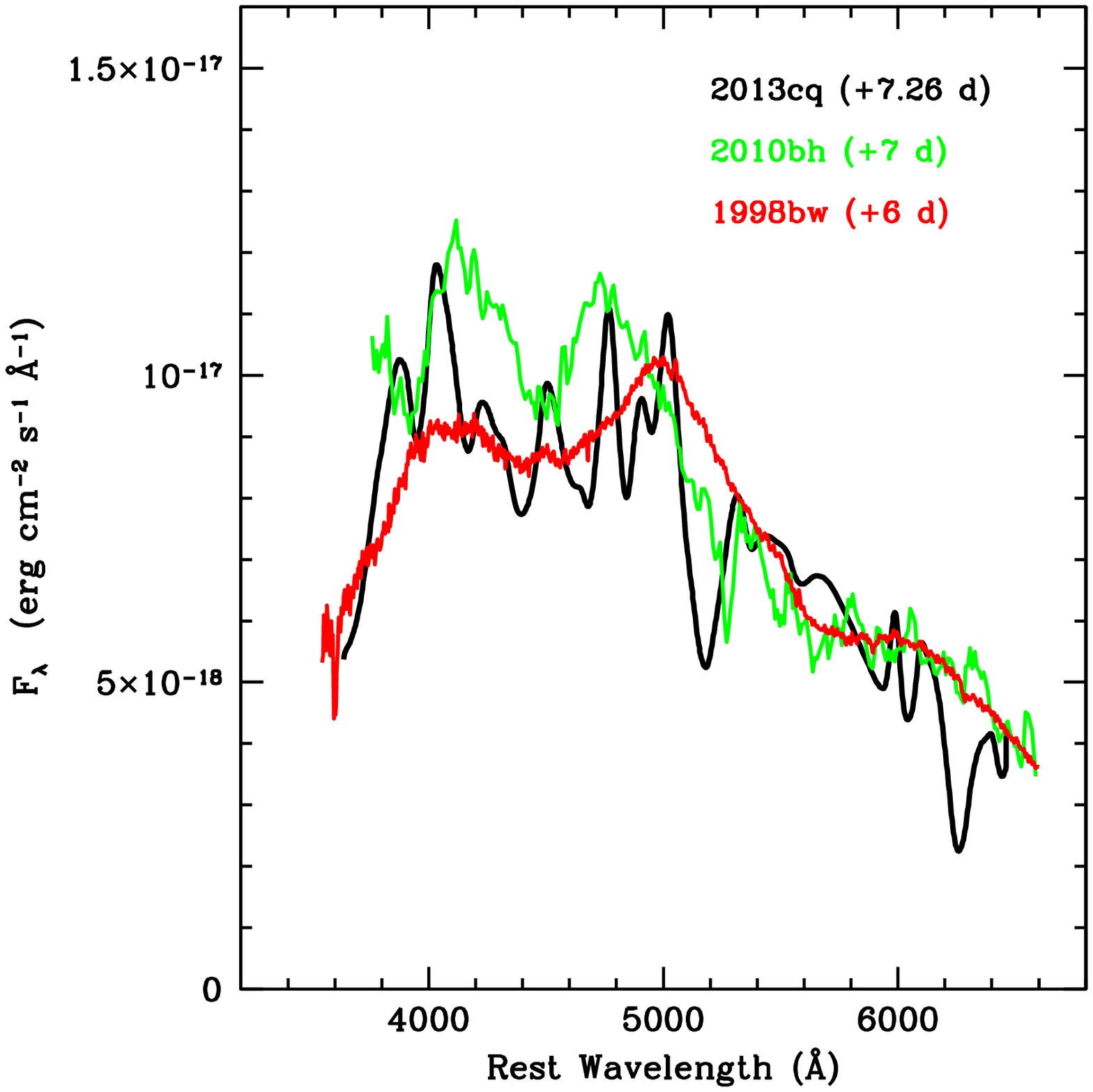}
      \vspace{-10mm}
            \hspace{-10mm}
    \includegraphics[width=10.5cm,height=6.7cm]{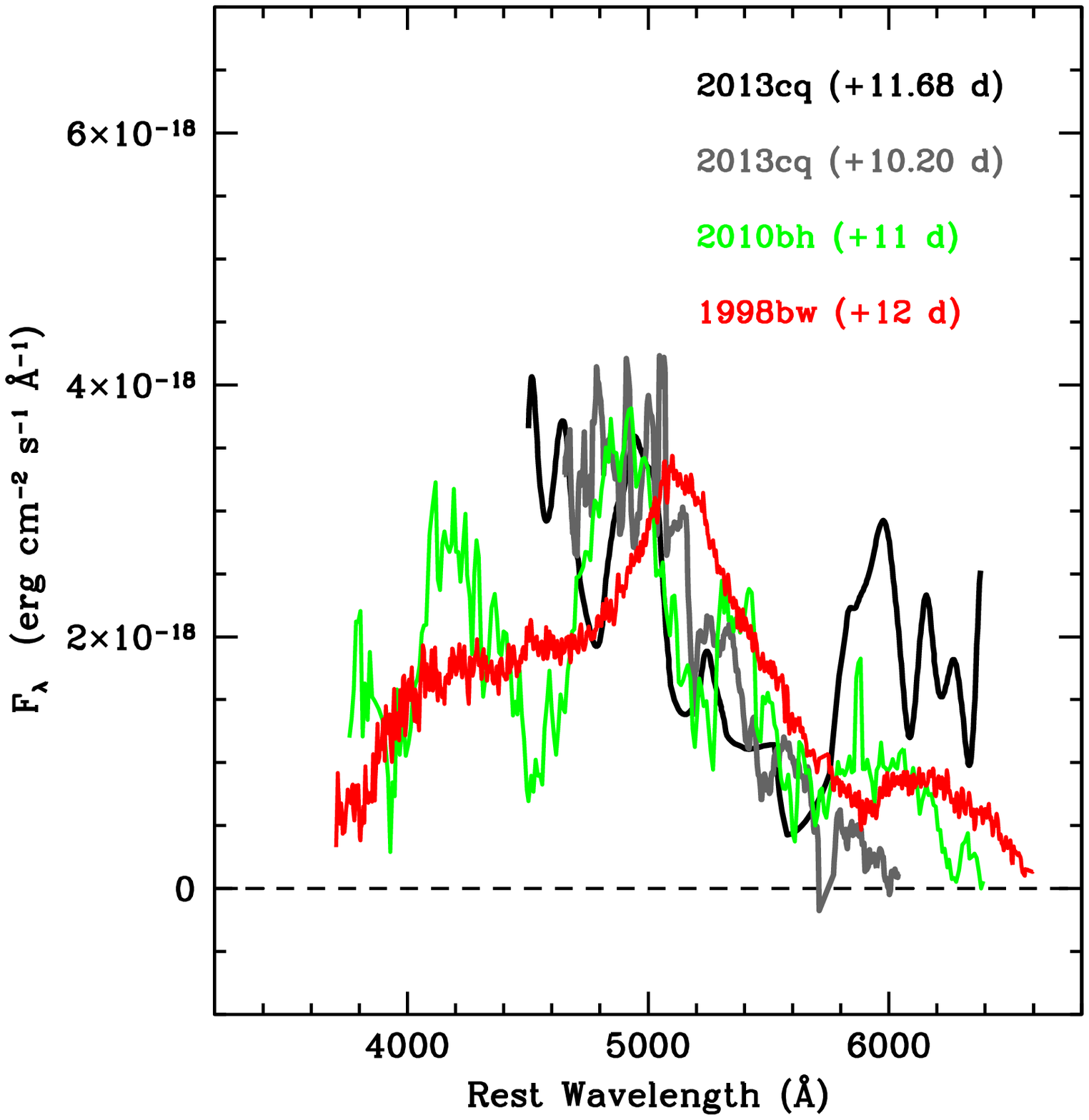}
     \vspace{3mm}
           \hspace{-8mm}
   \includegraphics[width=10.5cm,height=6.7cm]{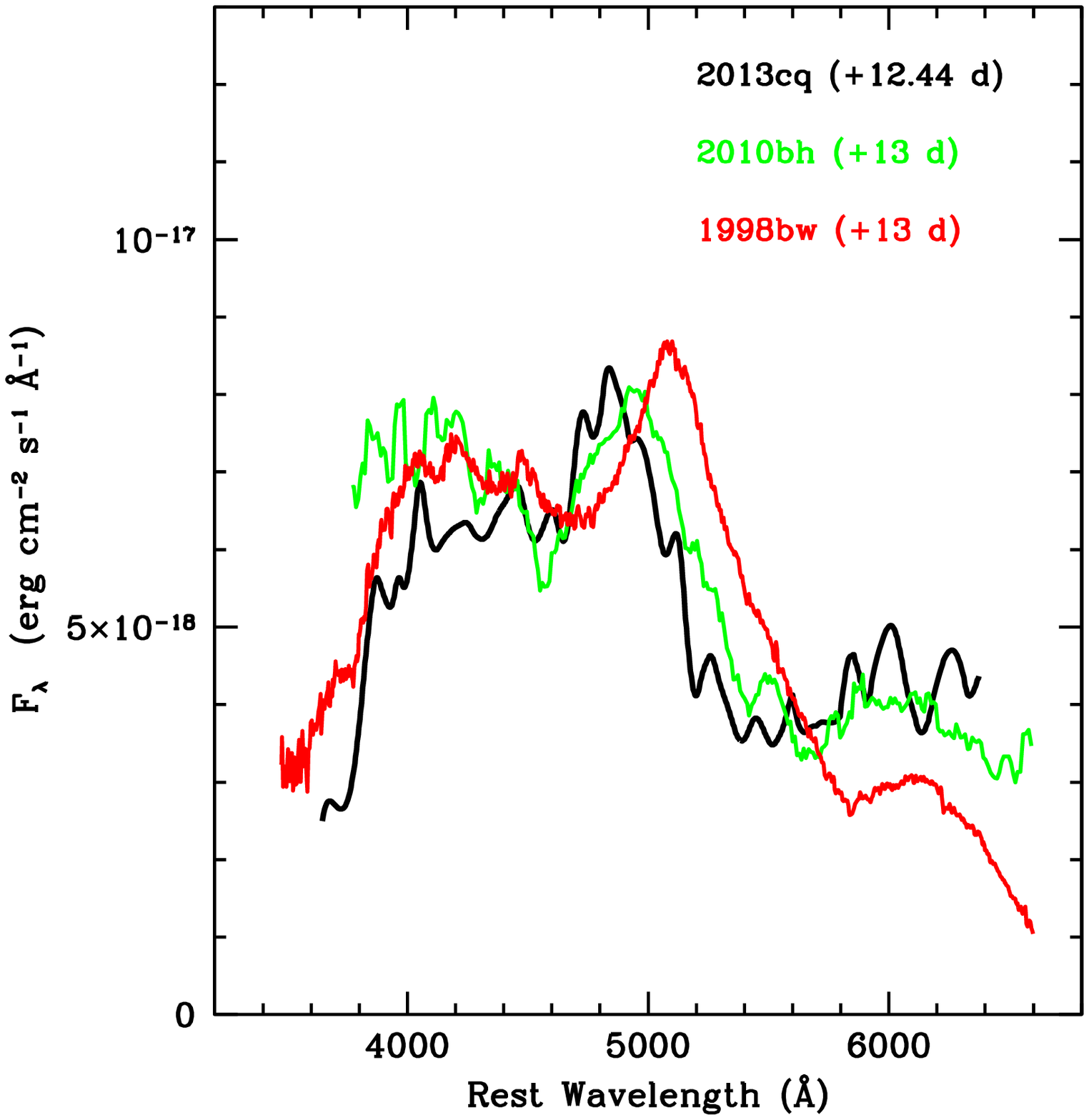}
     \vspace{5mm}
      \hspace{-10mm}
   \includegraphics[width=10.5cm,height=6.7cm]{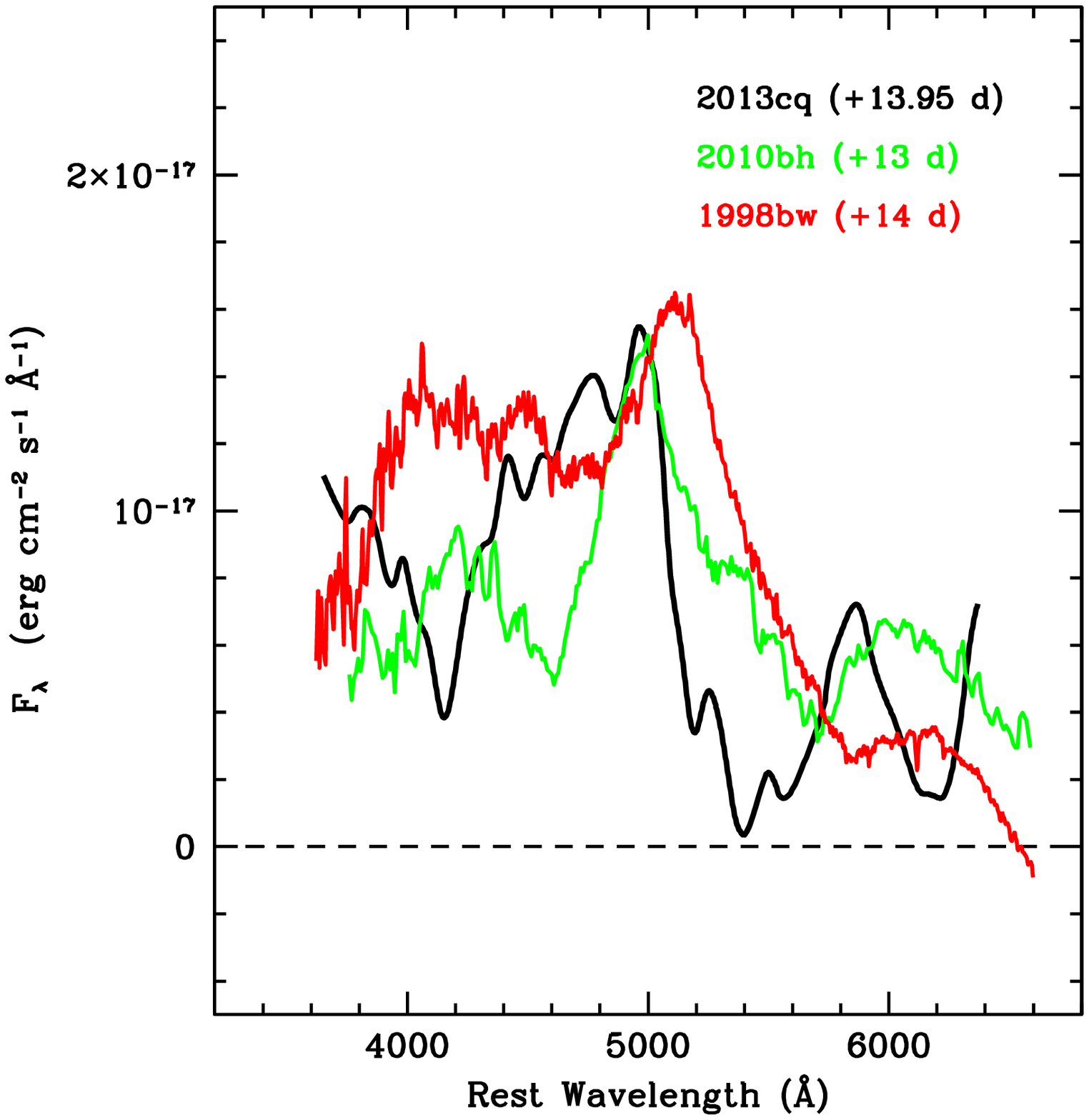}
      \vspace{-10mm}
            \hspace{-8mm}
   \includegraphics[width=10.5cm,height=6.7cm]{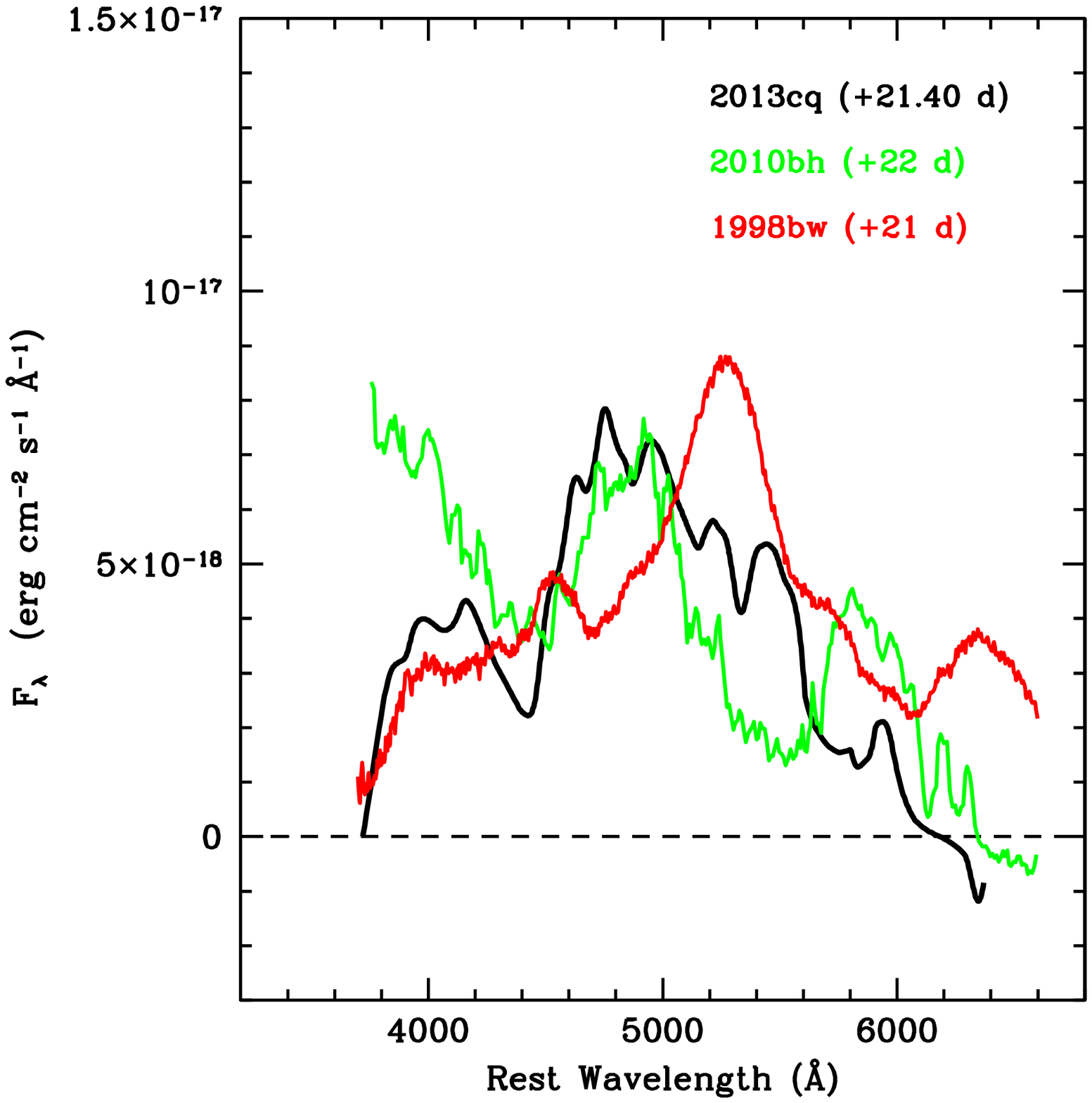}
      \vspace{-5mm}
            \hspace{-10mm}
    \includegraphics[width=10.5cm,height=6.7cm]{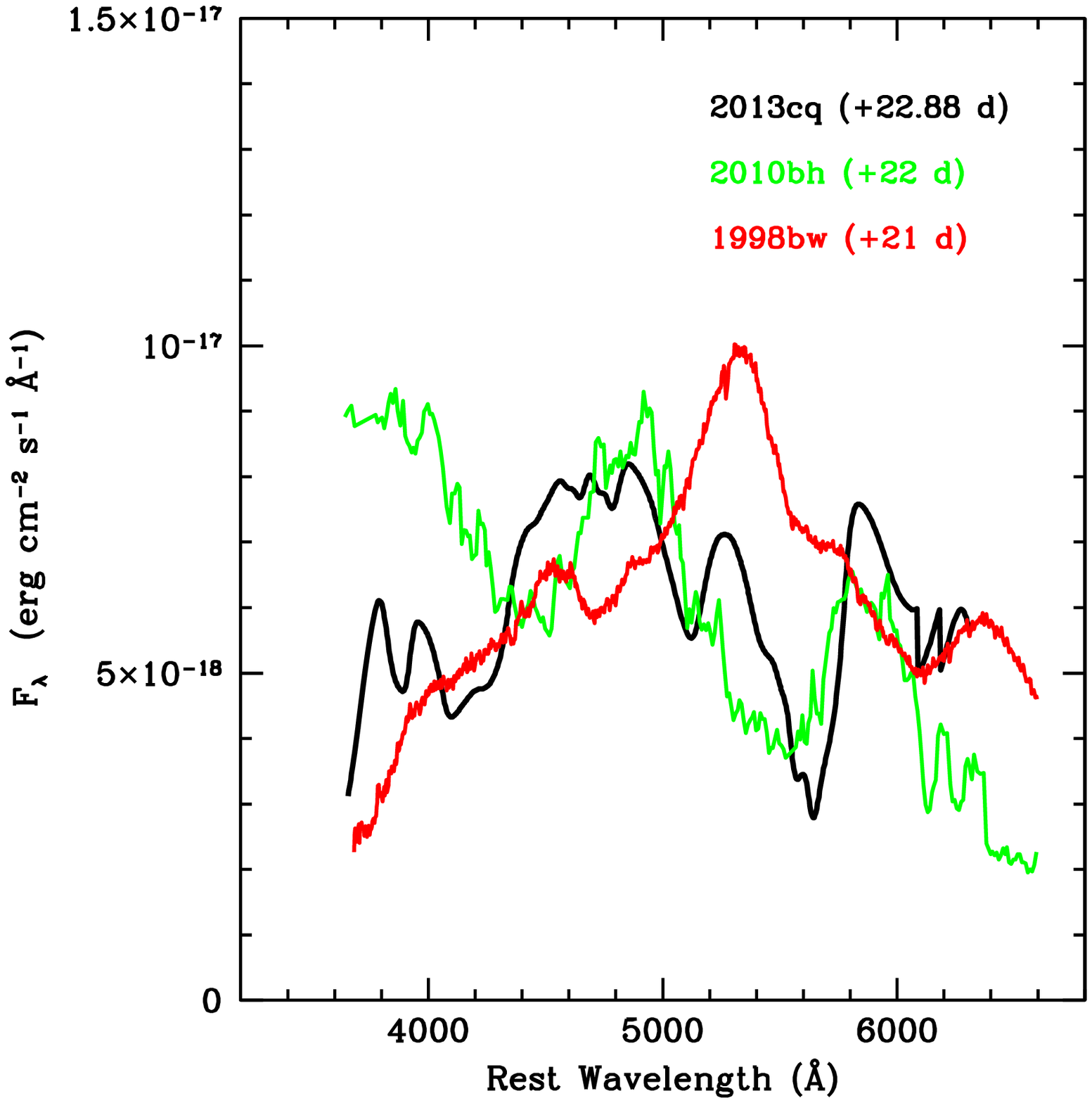}
       \vspace{-5mm}
             \hspace{-8mm}
   \includegraphics[width=10.5cm,height=6.7cm]{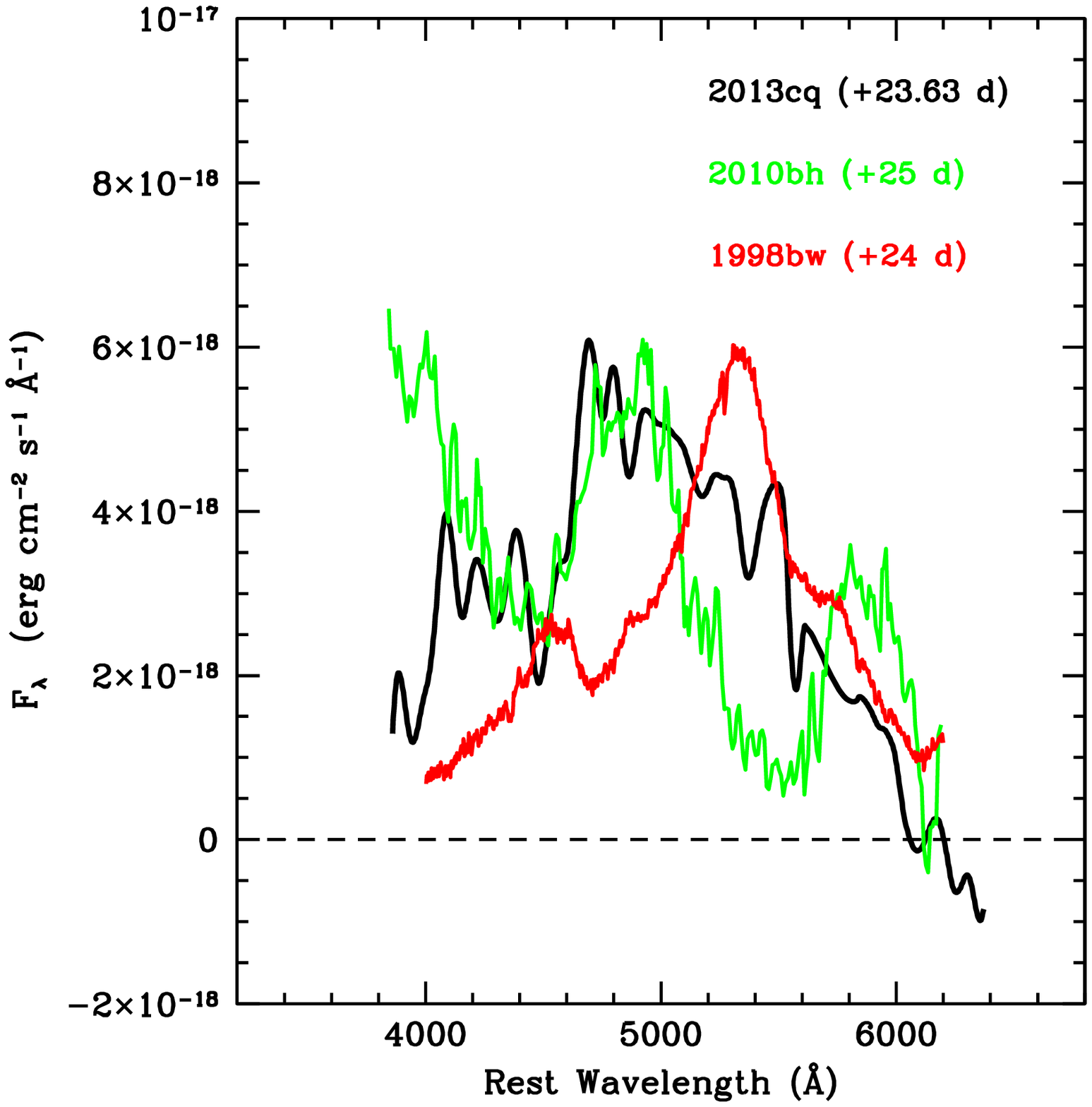}
       \vspace{5mm}
         \caption{Spectral evolution of SN\,2013cq: each panel shows the signal obtained with VLT-300V grism from SN\,2013cq (black) and the comparison with SN\,2010bh (green) and SN\,1998bw (red) at comparable rest-frame phases. We superpose, when available, the signal obtained with the VLT-300I grism (gray). All spectra have been smoothed with a boxcar of 15 $\AA$.}
    \label{spec2SN}
\end{figure*}

In Fig. \ref{spec2SN} we show the FORS2 spectra acquired between $\sim$ 5.0 and $\sim$ 23.6 rest-frame days after the burst event. 

The quality of our spectra is not sufficient for the accurate measurement of individual absorption features and for the estimate of photospheric velocities, but only for a general comparison with the spectra of high kinetic energy Type Ic SNe.  Among these we have selected some with good S/N and/or spectral coverage: the GRB SNe 1998bw (\cite{patat}) and 2003dh (\cite{hjorth}; \cite{mazzali}), XRF SNe 2006aj (\cite{pian}) and 2010bh (\cite{bufano}) and  the broad-lined Ic SN2010ah (\cite{mazzali3}). A  similarity is found with SN\,2010bh (see also \cite{xu2}) and with SN\,1998bw and for clarity we show only the comparison with these two SNe in Fig. \ref{spec2SN}. The spectra of SN\,2010bh are a good match of those of  SN\,2013cq, although those of  SN\,1998bw are a better representation in the blue, especially at phases +12.44, +21.40, +22.88 days, suggesting strong line-blocking from high velocity material.

\subsection{Bolometric light curve}

We have constructed a bolometric light curve in the range 3000-10000~$\AA$ using the available photometry. For each epoch we followed the reduction procedure described in section 3.

\begin{figure}
   \centering
\hspace{10mm}
   \includegraphics[width=9.8cm,height=9cm]{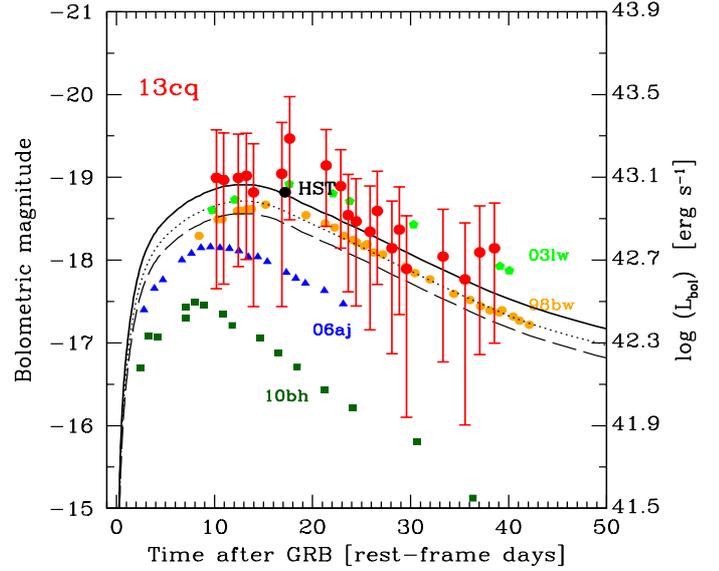}
      \caption{Bolometric light curve of SN\,2013cq (red filled circles) compared with those of SN\,1998bw (orange filled circles), SN\,2006aj (blue filled triangles), SN\,2010bh (green filled squares) and with the models for SN\,2003dh (dotted line) and SN\,2012bz (dashed line). We show also the best fit model for SN\,2013cq (solid line). As a consistency check, we also report the bolometric point (black filled circle) obtained from the HST measurements of May 20, 2013. The error on this point is within the size of the symbol.}
    \label{figbol}
\end{figure}

Then we fitted with a spline function the residual monochromatic light curves, which represent the supernova component, and integrated the broad-band flux at each photometric observation epoch. The flux was linearly extrapolated blueward of the $V$-band flux down to 3000~$\AA$ and redward of the $I$-band flux to 10000~$\AA$. The result is reported in Fig. \ref{figbol}. The errors associated with our photometry, galaxy measurement, afterglow fit and intrinsic absorption were propagated and summed in quadrature. These are extremely large at  the earlier epochs, so that the points have been  omitted in the figure. For comparison, and as a consistency check, we report  the bolometric point obtained from the HST measurements of May 20, 2013 (\cite{levan2}). We also compare the bolometric light curve of SN\,2013cq with those of SN\,1998bw (\cite{patat}), SN\,2006aj (\cite{pian}) and with the models for SN\,2003dh (\cite{mazzali4}) and SN\,2012bz (\cite{mela}).  The bolometric light curve of SN\,1998bw was constructed in the same rest-frame band (3000-10000~$\AA$) and using a Galactic extinction of $E_{B-V}$ = 0.052 mag as recently reported by Schlafly $\&$ Finkbeiner (2011).  It does not differ significantly from that reported in Pian et al. (2006) which was corrected for a lower estimate of Galactic extinction and included a $\sim$ 15\% flux correction for NIR contribution. The model of the light curve of SN\,2003dh (\cite{mazzali4}) was rescaled by 20\% to match the HST point, which is very accurate. The rescaled model also fits well the rest of the light curve, within the large errors.  This suggests that the $^{56}$Ni mass synthesized by SN\,2013cq is $\sim$20\% higher than that of SN\,2003dh, which leads to an estimate of $\sim 0.4 M_\odot$ (\cite{mazzali6}).

 \begin{table}
    \hspace{-10mm}
 \caption[]{GRBs-SNe properties. Columns are: (1) GRB-SN name, (2) redshift, (3) isotropic $\gamma$-ray energy in the 1-10000~keV energy band and (4) peak bolometric magnitude of the supernova.}
 \label{grbsn}
 \centering
 \begin{tabular}{@{\hspace{0pt}}c@{\hspace{6pt}}c@{\hspace{6pt}}c@{\hspace{6pt}}c@{\hspace{15pt}}l}
 \hline
GRB - SN & $z$ & $E_{\rm \gamma , iso}$ & $M_{\rm bol}$ & Refs.\\
\hline
 & & ($\times 10^{52}$~erg) & &\\
\hline
980425 - 1998bw & 0.0085 & 0.00010 $\pm$ 0.00002 & -18.65 $\pm$ 0.20 & 1,2\\
030329 - 2003dh & 0.1687 & 1.5 $\pm$ 0.3 & -18.71 $\pm$ 0.15 & 3,4\\
031203 - 2003lw & 0.1055 & 0.010 $\pm$ 0.004 & -18.92 $\pm$ 0.20 & 5,1\\
060218 - 2006aj & 0.0334 & 0.0053 $\pm$ 0.0003 & -18.16 $\pm$ 0.20 & 4,6,7\\
100316D - 2010bh & 0.0591 & 0.007 $\pm$ 0.003 & -17.50 $\pm$ 0.25 & 8,9\\
120422A - 2012bz & 0.283 & 0.024 $\pm$ 0.008 & -18.56 $\pm$ 0.15 & 10\\
130427A - 2013cq & 0.3399 & 81.0 $\pm$ 8.0 & -18.91 $\pm$ 0.20 & 11,2\\
130702A - 2013dx & 0.145 & 0.065 $\pm$ 0.010 & -18.51 $\pm$ 0.15 & 12,13\\
\hline
\multicolumn{5}{l}{\scriptsize{Refs: 1-Amati et al. (2006); 2-This work; 3-Deng et al. (2005); 4-Amati et al. (2008);}}\\
\multicolumn{5}{l}{\scriptsize{5-Mazzali et al. (2006a); 6-Pian et al. (2006); 7-Ferrero et al. (2006);}}\\
\multicolumn{5}{l}{\scriptsize{8-Bufano et al. (2012); 9-Starling et al. (2011); 10-Melandri et al. (2012);}}\\
\multicolumn{5}{l}{\scriptsize{11-Maselli et al. 2014; 12-D'Elia et al. (2014, in preparation);}}\\
\multicolumn{5}{l}{\scriptsize{13-Amati et al. (2013b; note that uncertainty on $E_{\rm \gamma , iso}$ is amended here).}}\\
\end{tabular}
\end{table}

\subsection{GRBs-SNe properties and correlations}

GRB\,130427A is the brightest GRB detected by {\it Swift}/BAT, one of the most energetic ever ($E_{\rm \gamma , iso} \sim 10^{54}$~erg, $E_{\rm \gamma , peak} \sim 1.2-1.3 \times 10^{3}$~keV), and the most energetic GeV emitting GRB ( $\sim$125 GeV in rest-frame; \cite{zhu2}). GRB\,130427A is located in a yet unexplored region of the $E_{\rm \gamma , iso}$-$E_{\rm \gamma , peak}$ plane for low-$z$ GRBs associated with SNe, yet it follows very well the Amati and Yonetoku correlations (see Fig. S6 in \cite{maselli2}). Prior to SN\,2013cq only one nearby supernova was associated with a GRB having energetics similar to cosmological gamma-ray bursts, i.e. GRB\,030329 ($E_{\rm \gamma , iso} \sim 2 \times 10^{52}$~erg, $E_{\rm \gamma , peak} = 82 \pm 3$~keV; \cite{spek}). However, the properties of the GRB\,030329 were less extreme than those observed for GRB\,130427A.

\begin{figure}
   \centering
      \includegraphics[width=8.cm,height=8.0cm]{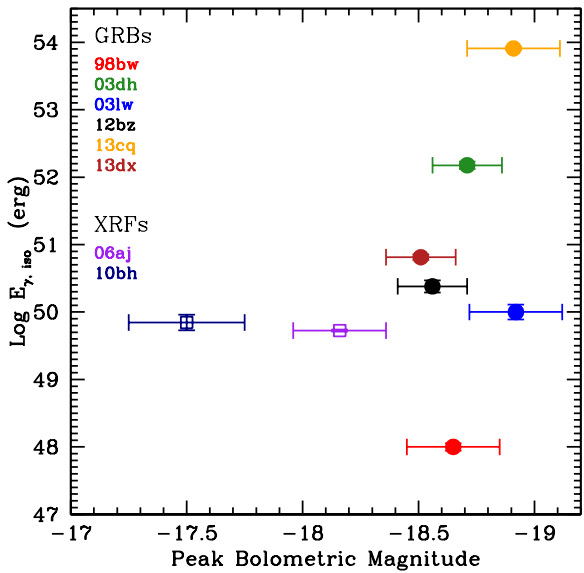}
   \caption{GRB $E_{\rm \gamma , iso}$ versus the peak magnitude of the bolometric light curve of the corresponding SN: the lack of correlation is apparent, being the SN brightness nearly constant while the GRB energy spans several orders of magnitude. }
    \label{correla}
\end{figure}

A striking result when comparing GRB and SN properties (see Table 2) is that, while the values of E$_{\rm iso}$ of the GRB span nearly 6 orders of magnitude ($\sim$3 after correcting for collimation effects) the SNe maximum luminosities ($M_{\rm bol}$), that trace the mass of radioactive $^{56}$Ni and correlate, like the SN kinetic energies, with the progenitor masses (\cite{mazzali3}), are distributed in a narrow range ($\sim 0.5$~mag) and can virtually be considered roughly constant. On the other hand, nearby XRFs have total isotropic-equivalent energies similar to those of the less energetic GRBs, but their SNe have lower luminosities (Fig. \ref{correla}).

\section{Conclusion}

The $R$-band light curve reported by Xu et al. (2013b, see also \cite{perley}) is $\sim$ 0.2 mag fainter than that of SN\,1998bw at maximum (\cite{patat}). Our analysis seems to confirm this (Fig. \ref{lc_rf_SN}).  However, the bolometric light curve (Fig. \ref{figbol}), and especially the accurate HST measurement (\cite{levan2}), suggest that SN\,2013cq is marginally brighter than SN\,1998bw and significantly more luminous ($\sim 1.5$~mag) than SN\,2010bh. This points to the need of referring to bolometric rather than monochromatic information in the comparison of SN physical quantities. 

Although the S/N of the spectra is very limited, we note some similarity with SN\,2010bh (see also \cite{xu2}) and with SN\,1998bw. In either case, this fact points to a classification of SN\,2013cq as a broad line type Ic SN, as seen so far for all GRB/XRF SNe, and in turn to a massive, highly envelope-stripped progenitor.

This is in line with the finding that SNe associated with GRBs -- both underluminous and highly energetic -- have all comparable luminosities and are more luminous than the two known low-$z$ spectroscopically identified XRF/SNe 2006aj and 2010bh (\cite{mela}). The extended host galaxy of GRB\,130427A/SN\,2013cq is similar to the host galaxies of GRB\,100316D/SN\,2010bh and GRB\,980425/SN\,1998bw. The other ones known at low redshift are instead associated with smaller galaxies (\cite{starling}). However there is no apparent correlation between the brightness of the GRB, the associated SNe and their host galaxies (\cite{levesque}).

Owing to the limited S/N of the spectra of SN\,2013cq, it is difficult to evaluate the photospheric velocity evolution of this SN and therefore its kinetic energy. Xu et al. (2013b) estimate $\sim 6 \times 10^{52}$~erg, i.e. similar to SN\,1998bw, which is plausible, considering the similarity of spectra and bolometric light curve. The data reported by Maselli et al. (2013b, their Fig. 2) show the presence of a break in the afterglow light curve, which they interpret as a jet break. If this is the case (jet collimation-correction can be complex, \cite{campana}; \cite{perley}), after correcting for the corresponding jet opening angle of about 3\degr, the total energy of GRB\,130427A decreases to 3 $\times 10^{50}$~erg, which is less than 1$\%$ of the kinetic energy associated with its SN. A similar finding has been reported for other GRB-SN associations (\cite{wb}, \cite{amati4}). Therefore it is very plausible that the SN may drive the GRB jet. The monitoring of nearby energetic GRBs is a critical test-bed for confirming SN-driven GRB models or opening more exotic (and more energetic) scenarios based on black hole formation.

\begin{acknowledgements}

We thank the anonymous referee for valuable comments and suggestions that improved the paper. We thank the TNG staff, in particular G. Andreuzzi, L. Di Fabrizio, and M. Pedani, for their valuable support with TNG observations, and the Paranal Science Operations Team, in particular H. Boffin, S. Brillant, C. Cid, O. Gonzales, V. D. Ivanov, D. Jones, J. Pritchard, M. Rodrigues, L. Schmidtobreick, F. J. Selman, J. Smoker and S. Vega. The Dark Cosmology Centre is funded by the Danish National Research Foundation. F.B. acknowledges support from FONDECYT through Postdoctoral grant 3120227 and from Project IC120009 "Millennium Institute of Astrophysics (MAS)" of  the Iniciativa Cient'fica Milenio del Ministerio de Econom\'{i}a, Fomento y Turismo de Chile. A.J.C.T. thanks the Spanish Ministry's Research Project AYA 2012-39727-C03-01. R.L.C.S. is supported by a Royal Society fellowship. D.M. acknowledges the Instrument Center for Danish Astrophysics (IDA) for support. This research was partially supported by contracts ASI INAF I/004/11/1, ASI INAF I/088/06/0, INAF PRIN 2011 and PRIN MIUR 2010/2011. 

\end{acknowledgements}

%
%

\newpage

\onecolumn
 \begin{longtable}{lcccccc}
 \caption[]{Observations log: the first column shows the date of the observation, $\Delta t$ (column 2) refers to the beginning of the exposure, with respect to the GRB trigger time, with a duration $t_{\rm exp}$ (column 3). Magnitudes (column 4) have not been corrected for Galactic absorption along the line of sight ($E_{B-V}$ = 0.02 mag, \cite{schle}; \cite{schla}). We calculated flux densities (column 5) from absorption-corrected magnitudes, following Fukugita et al. 1995.} 
 \label{logmag}
\endfirsthead
\multicolumn{7}{r}{\textit{Continue to the next page}}
\endfoot
\endhead
\endlastfoot
Date & $\Delta t$ & $t_{\rm exp}$ & Magnitude & Flux density & Ref. & Seeing\\
\hline
 & (days) & (min) & & ($\mu$Jy) & & (\arcsec)\\
\hline
 2013 May 3    & 6.722   & 1.0 & 20.93  $\pm$ 0.02 & 19.4  $\pm$ 0.4 & VLT-$B$  & 0.96\\
 2013 May 6    & 9.723   & 1.0 & 21.37  $\pm$ 0.04 & 12.9  $\pm$ 0.5 & VLT-$B$  & 1.18\\
 2013 May 8    & 11.645  & 1.0 & 21.40  $\pm$ 0.04 & 12.5  $\pm$ 0.5 & VLT-$B$  & 0.77\\
 2013 May 10   & 13.648  & 1.0 & 21.58  $\pm$ 0.06 & 10.7  $\pm$ 0.6 & VLT-$B$  & 1.20\\
 2013 May 11   & 14.648  & 1.0 & 21.63  $\pm$ 0.12 & 10.2  $\pm$ 1.1 & VLT-$B$  & 1.57\\
 2013 May 13   & 16.644  & 1.5 & 21.74  $\pm$ 0.06 & 9.3  $\pm$ 0.5 & VLT-$B$  & 1.49\\
 2013 May 14   & 17.748  & 1.5 & 21.68  $\pm$ 0.04 & 9.7  $\pm$ 0.4 & VLT-$B$  & 0.80\\
 2013 May 15   & 18.720  & 1.5 & 21.75  $\pm$ 0.08 & 9.2  $\pm$ 0.7 & VLT-$B$  & 1.34\\
 2013 May 19   & 22.597  & 2.5 & 21.68  $\pm$ 0.08 & 9.7  $\pm$ 0.7 & TNG-$g'$ & 1.25\\
 2013 May 20   & 23.674  & 3.0 & 21.82  $\pm$ 0.22 & 8.6  $\pm$ 1.8 & VLT-$B$  & 1.00\\
 2013 May 25   & 28.657  & 1.5 & 22.11  $\pm$ 0.19 & 6.6  $\pm$ 1.2 & VLT-$B$  & 1.11\\
 2013 May 27   & 30.643  & 1.5 & 22.13  $\pm$ 0.07 & 6.4  $\pm$ 0.5 & VLT-$B$  & 1.38\\
 2013 May 28   & 31.647  & 1.5 & 22.22  $\pm$ 0.04 & 5.9  $\pm$ 0.2 & VLT-$B$  & 0.70\\
 2013 May 29   & 32.637  & 1.5 & 22.26  $\pm$ 0.07 & 5.7  $\pm$ 0.4 & VLT-$B$  & 0.96\\
 2013 June 1   & 34.649  & 1.5 & 22.25  $\pm$ 0.14 & 5.7  $\pm$ 0.7 & VLT-$B$  & 0.75\\
 2013 June 2   & 35.649  & 1.5 & 22.21  $\pm$ 0.06 & 6.0  $\pm$ 0.3 & VLT-$B$  & 0.97\\
 2013 June 3   & 37.649  & 1.5 & 22.20  $\pm$ 0.06 & 6.0  $\pm$ 0.4 & VLT-$B$  & 0.76\\
 2013 June 4   & 38.655  & 1.5 & 22.22  $\pm$ 0.08 & 5.9  $\pm$ 0.4 & VLT-$B$  & 0.72\\
 2013 June 5   & 39.649  & 1.5 & 22.22  $\pm$ 0.09 & 5.9  $\pm$ 0.5 & VLT-$B$  & 0.93\\
 2013 June 6   & 40.639  & 2.5 & 22.19  $\pm$ 0.08 & 6.1  $\pm$ 0.5 & VLT-$B$  & 0.81\\
 2013 June 10  & 44.639  & 1.5 & 22.21  $\pm$ 0.09 & 6.0  $\pm$ 0.5 & VLT-$B$  & 0.98\\
 2013 June 13  & 47.626  & 3.0 & 22.21  $\pm$ 0.23 & 6.0  $\pm$ 1.3 & VLT-$B$  & 1.29\\
 2013 June 15  & 49.649  & 1.5 & 22.24  $\pm$ 0.21 & 5.8  $\pm$ 1.1 & VLT-$B$  & 1.57\\
 2013 June 16  & 50.671  & 1.5 & 22.20  $\pm$ 0.18 & 6.1  $\pm$ 1.0 & VLT-$B$  & 1.25\\
 2013 June 17  & 51.670  & 1.5 & 22.19  $\pm$ 0.22 & 6.1  $\pm$ 1.2 & VLT-$B$  & 0.96\\
\hline
 2013 May 3    & 6.724   & 1.0  & 20.31   $\pm$ 0.02  & 28.7  $\pm$ 0.6  & VLT-$V$  & 1.16\\
 2013 May 6    & 9.724   & 1.0  & 20.60   $\pm$ 0.03  & 22.2  $\pm$ 0.6  & VLT-$V$  & 1.20\\
 2013 May 8    & 11.647  & 1.0  & 20.74   $\pm$ 0.02  & 19.4  $\pm$ 0.5  & VLT-$V$  & 0.85\\
 2013 May 10   & 13.650  & 1.0  & 20.81   $\pm$ 0.04  & 18.2  $\pm$ 0.7  & VLT-$V$  & 1.37\\
 2013 May 11   & 14.649  & 1.0  & 20.84   $\pm$ 0.03  & 17.8  $\pm$ 0.4  & VLT-$V$  & 1.15\\
 2013 May 13   & 16.646  & 1.5  & 20.90   $\pm$ 0.03  & 16.8  $\pm$ 0.5  & VLT-$V$  & 1.30\\
 2013 May 14   & 17.751  & 1.5  & 20.96   $\pm$ 0.06  & 15.9  $\pm$ 0.8  & VLT-$V$  & 0.89\\
 2013 May 15   & 18.723  & 1.5  & 20.99   $\pm$ 0.04  & 15.4  $\pm$ 0.6  & VLT-$V$  & 1.34\\
 2013 May 20   & 23.677  & 3.0  & 20.85   $\pm$ 0.13  & 17.7  $\pm$ 2.2  & VLT-$V$  & 1.17\\
 2013 May 25   & 28.657  & 1.5  & 21.05   $\pm$ 0.08  & 14.7  $\pm$ 1.1  & VLT-$V$  & 1.07\\
 2013 May 27   & 30.645  & 1.5  & 21.18   $\pm$ 0.05  & 13.0  $\pm$ 0.5  & VLT-$V$  & 1.25\\
 2013 May 28   & 31.649  & 1.5  & 21.25   $\pm$ 0.03  & 12.2  $\pm$ 0.3  & VLT-$V$  & 0.68\\
 2013 May 29   & 32.639  & 1.5  & 21.25   $\pm$ 0.04  & 12.2  $\pm$ 0.4  & VLT-$V$  & 1.03\\
 2013 June 1   & 34.651  & 1.5  & 21.33   $\pm$ 0.06  & 11.3  $\pm$ 0.7  & VLT-$V$  & 0.76\\
 2013 June 2   & 35.652  & 1.5  & 21.34   $\pm$ 0.04  & 11.2  $\pm$ 0.4  & VLT-$V$  & 1.33\\
 2013 June 3   & 37.651  & 1.5  & 21.33   $\pm$ 0.03  & 11.3  $\pm$ 0.3  & VLT-$V$  & 0.75\\
 2013 June 4   & 38.657  & 1.5  & 21.31   $\pm$ 0.03  & 11.6  $\pm$ 0.4  & VLT-$V$  & 0.91\\
 2013 June 5   & 39.652  & 1.5  & 21.44   $\pm$ 0.04  & 10.3  $\pm$ 0.4  & VLT-$V$  & 0.80\\
 2013 June 6   & 40.643  & 2.5  & 21.37   $\pm$ 0.03  & 10.9  $\pm$ 0.3  & VLT-$V$  & 0.86\\
 2013 June 10  & 44.641  & 1.5  & 21.41   $\pm$ 0.04  & 10.5  $\pm$ 0.4  & VLT-$V$  & 0.92\\
 2013 June 13  & 47.628  & 3.0  & 21.48   $\pm$ 0.08  & 9.9  $\pm$ 0.7  & VLT-$V$  & 1.62\\
 2013 June 15  & 49.652  & 1.5  & 21.38   $\pm$ 0.06  & 10.8  $\pm$ 0.6  & VLT-$V$  & 1.62\\
 2013 June 16  & 50.673  & 1.5  & 21.57   $\pm$ 0.08  & 9.1  $\pm$ 0.7  & VLT-$V$  & 0.85\\
 2013 June 17  & 51.660  & 1.5  & 21.41   $\pm$ 0.08  & 10.5  $\pm$ 0.8  & VLT-$V$  & 0.87\\
   \hline
 2013 April 30 &  3.651  & 1.0 & 19.43  $\pm$  0.06 &  54.8  $\pm$  3.3 & TNG-$r'$ & 1.56 \\
 2013 May 4    &  6.726  & 1.0 & 20.07  $\pm$  0.05 &  30.3  $\pm$  1.4 & VLT-$R$ & 0.88 \\
 2013 May 7    &  9.727  & 1.0 & 20.35  $\pm$  0.03 &  23.4  $\pm$  0.6 & VLT-$R$ & 1.30 \\
 2013 May 8    &  11.649 & 1.0 & 20.38  $\pm$  0.03 &  22.8  $\pm$  0.6 & VLT-$R$ & 1.03 \\
 2013 May 10   &  13.652 & 1.0 & 20.47  $\pm$  0.04 &  21.0  $\pm$  0.8 & VLT-$R$ & 1.40 \\
 2013 May 11   &  14.657 & 1.0 & 20.52  $\pm$  0.04 &  20.0  $\pm$  0.7 & VLT-$R$ & 1.21 \\
 2013 May 13   &  16.650 & 1.5 & 20.57  $\pm$  0.03 &  19.1  $\pm$  0.5 & VLT-$R$ & 1.12 \\
 2013 May 14   &  17.755 & 1.5 & 20.59  $\pm$  0.04 &  18.7  $\pm$  0.6 & VLT-$R$ & 0.86 \\
 2013 May 15   &  18.726 & 1.5 & 20.50  $\pm$  0.09 &  20.4  $\pm$  1.7 & VLT-$R$ & 1.34 \\
 2013 May 19   &  22.607 & 1.0 & 20.69  $\pm$  0.16 &  17.1  $\pm$  2.7 & TNG-$r'$ & 1.40 \\
 2013 May 20   &  23.680 & 3.0 & 20.41  $\pm$  0.12 &  22.2  $\pm$  2.5 & VLT-$R$ & 1.17 \\
 2013 May 25   &  28.660 & 1.5 & 20.72  $\pm$  0.05 &  16.7  $\pm$  0.8 & VLT-$R$ & 0.90 \\
 2013 May 27   &  30.648 & 1.5 & 20.81  $\pm$  0.04 &  15.4  $\pm$  0.6 & VLT-$R$ & 1.22 \\
 2013 May 28   &  31.654 & 1.5 & 20.88  $\pm$  0.02 &  14.4  $\pm$  0.3 & VLT-$R$ & 0.77 \\
 2013 May 29   &  32.757 & 1.5 & 20.20  $\pm$  0.03 &  14.2  $\pm$  0.4 & VLT-$R$ & 0.97 \\
 2013 June 1   &  34.655 & 1.5 & 20.91  $\pm$  0.06 &  14.0  $\pm$  0.7 & VLT-$R$ & 0.77 \\
 2013 June 2   &  35.656 & 1.5 & 20.92  $\pm$  0.04 &  13.9  $\pm$  0.6 & VLT-$R$ & 1.18 \\
 2013 June 3   &  37.654 & 1.5 & 20.88  $\pm$  0.02 &  14.3  $\pm$  0.3 & VLT-$R$ & 0.70 \\
 2013 June 4   &  38.660 & 1.5 & 20.94  $\pm$  0.03 &  13.6  $\pm$  0.4 & VLT-$R$ & 0.87 \\
 2013 June 5   &  39.651 & 1.5 & 20.94  $\pm$  0.03 &  13.6  $\pm$  0.3 & VLT-$R$ & 0.79 \\
 2013 June 6   &  40.642 & 2.5 & 20.95  $\pm$  0.02 &  13.5  $\pm$  0.3 & VLT-$R$ & 0.89 \\
 2013 June 10  &  44.640 & 1.5 & 21.09  $\pm$  0.04 &  11.8  $\pm$  0.5 & VLT-$R$ & 1.00 \\
 2013 June 13  &  47.627 & 3.0 & 21.05  $\pm$  0.06 &  12.3  $\pm$  0.7 & VLT-$R$ & 1.34 \\
 2013 June 15  &  49.651 & 1.5 & 21.16  $\pm$  0.09 &  11.1  $\pm$  0.9 & VLT-$R$ & 1.56 \\
 2013 June 16  &  50.672 & 1.5 & 21.10  $\pm$  0.07 &  11.7  $\pm$  0.8 & VLT-$R$ & 1.24\\
 2013 June 17  &  51.659 & 1.5 & 21.24  $\pm$  0.05 &  10.3  $\pm$  0.5 & VLT-$R$ & 0.90\\
\hline
 2013 April 30 &  3.652  & 1.0 & 18.95  $\pm$  0.06 &  65.7  $\pm$  3.9 & TNG-$i'$ & 1.34\\
 2013 May 3    &  6.727  & 1.0 & 19.40  $\pm$  0.02 &  43.6  $\pm$  0.9 & VLT-$I$  & 0.91\\
 2013 May 6    &  9.728  & 1.0 & 19.68  $\pm$  0.03 &  33.7  $\pm$  1.0 & VLT-$I$  & 1.14\\
 2013 May 8    &  11.650 & 1.0 & 19.85  $\pm$  0.03 &  28.8  $\pm$  0.8 & VLT-$I$  & 0.79\\
 2013 May 10   &  13.653 & 1.0 & 19.85  $\pm$  0.04 &  28.7  $\pm$  1.1 & VLT-$I$  & 1.40\\
 2013 May 11   &  14.658 & 1.0 & 19.85  $\pm$  0.03 &  28.9  $\pm$  0.9 & VLT-$I$  & 0.73\\
 2013 May 13   &  16.651 & 1.5 & 19.90  $\pm$  0.05 &  27.4  $\pm$  1.3 & VLT-$I$  & 0.89\\
 2013 May 14   &  17.756 & 1.5 & 19.88  $\pm$  0.06 &  28.0  $\pm$  1.7 & VLT-$I$  & 0.94\\
 2013 May 15   &  18.727 & 1.5 & 20.07  $\pm$  0.13 &  24.7  $\pm$  2.6 & VLT-$I$  & 1.34\\
 2013 May 19  &  22.593 & 1.0 & 20.02  $\pm$  0.16 &  24.6  $\pm$  3.6 & TNG-$i'$ & 1.16\\
 2013 May 20   &  23.681 & 3.0 & 19.83  $\pm$  0.16 &  29.3  $\pm$  4.5 & VLT-$I$  & 1.17\\
 2013 May 25   &  28.661 & 1.5 & 19.88  $\pm$  0.03 &  28.0  $\pm$  0.9 & VLT-$I$  & 0.92\\
 2013 May 27   &  30.649 & 1.5 & 19.93  $\pm$  0.04 &  26.7  $\pm$  1.1 & VLT-$I$  & 1.46\\
 2013 May 28   &  31.655 & 1.5 & 20.07  $\pm$  0.03 &  23.4  $\pm$  0.7 & VLT-$I$  & 0.86\\
 2013 May 29   &  32.758 & 1.5 & 20.13  $\pm$  0.04 &  22.3  $\pm$  0.9 & VLT-$I$  & 1.00\\
 2013 June 1   &  34.656 & 1.5 & 20.13  $\pm$  0.05 &  22.2  $\pm$  0.9 & VLT-$I$  & 0.71\\
 2013 June 2   &  35.657 & 1.5 & 19.96  $\pm$  0.04 &  26.0  $\pm$  1.1 & VLT-$I$  & 1.25\\
 2013 June 3   &  37.655 & 1.5 & 20.30  $\pm$  0.03 &  19.1  $\pm$  0.6 & VLT-$I$  & 0.69\\
 2013 June 4   &  38.661 & 1.5 & 20.18  $\pm$  0.04 &  21.2  $\pm$  0.8 & VLT-$I$  & 0.96\\
 2013 June 5   &  39.652 & 1.5 & 20.27  $\pm$  0.03 &  19.5  $\pm$  0.6 & VLT-$I$  & 0.76\\
 2013 June 6   &  40.643 & 2.5 & 20.13  $\pm$  0.03 &  22.2  $\pm$  0.7 & VLT-$I$  & 0.91\\
 2013 June 10  &  44.641 & 1.5 & 20.25  $\pm$  0.04 &  19.9  $\pm$  0.8 & VLT-$I$  & 0.88\\
 2013 June 13  &  47.628 & 3.0 & 20.30  $\pm$  0.03 &  18.9  $\pm$  0.5 & VLT-$I$  & 1.34\\
 2013 June 15  &  49.652 & 1.5 & 20.33  $\pm$  0.04 &  18.5  $\pm$  0.7 & VLT-$I$  & 1.49\\
 2013 June 16  &  50.673 & 1.5 & 20.35  $\pm$  0.04 &  18.2  $\pm$  0.6 & VLT-$I$  & 0.94\\
 2013 June 17  &  51.660 & 1.5 & 20.27  $\pm$  0.03 &  19.5  $\pm$  0.6 & VLT-$I$  & 1.09\\
\end{longtable}
\twocolumn

\end{document}